\documentstyle[12pt,aaspp4,epsf]{article}		

\tighten		

\slugcomment{2001, ApJ, 547, 1 February}

\begin{document}

\title{Ultra-Lithium-Deficient Halo Stars and Blue Stragglers: 
A Common Origin?\footnote{Based on observations obtained with 
the University College London \'echelle spectrograph (UCLES) on the
Anglo-Australian Telescope (AAT) and the Utrecht \'echelle spectrograph (UES)
on the William Herschel Telescope (WHT).}}

\author{Sean G. Ryan}
\affil{Dept of Physics and Astronomy, The Open University, Walton Hall, 
MK7 6AA, UK.
email: s.g.ryan@open.ac.uk}

\author{Timothy C. Beers}
\affil{Dept of Physics and Astronomy, Michigan State University, East Lansing,
MI 48824, USA. email: beers@pa.msu.edu}

\author{Toshitaka Kajino}
\affil{National Astronomical Observatory of Japan, Osawa 2-21-1, Mitaka, 
Tokyo 181-8588, Japan, 
and 
Department of Astronomy, University of Tokyo, 7-3-1 Hongo, Bunkyo-ku, 
Tokyo 113-0015, Japan.
email: kajino@ferio.mtk.nao.ac.jp}

\and

\author{Katarina Rosolankova}
\affil{Dept of Physics and Astronomy, The Open University, Walton Hall, 
Milton Keynes MK7 6AA, UK.
email: katarina.rosolankova@st-hildas.oxford.ac.uk}

\begin{abstract}

We present data for four ultra-Li-deficient, warm, halo stars. The Li
deficiency of two of these is a new discovery.  Three of the four stars
have effective temperatures $T_{\rm eff}~\sim~6300$~K, in contrast to
previously known Li-deficient halo stars which spanned the temperature range of
the Spite Plateau.  In this paper we propose that these, and previously known
ultra-Li-deficient halo stars, may have had their surface lithium abundances
reduced by the same mechanism as produces halo field blue stragglers.  Even
though these stars have yet to reveal themselves as blue stragglers, they might
be regarded as ``blue-stragglers-to-be.'' In our proposed scenario, the surface
abundance of Li in these stars could be destroyed (a) during the normal
pre-main-sequence single star evolution of their low mass precursors, (b)
during the post-main-sequence evolution of a evolved mass donor, and/or (c) via
mixing during a mass-transfer event or stellar merger.  The warmest
Li-deficient stars at the turnoff would be regarded as emerging ``canonical''
blue stragglers, whereas cooler ones represent sub-turnoff-mass
``blue-stragglers-to-be.''  The latter are presently hidden on the main
sequence, Li depletion being possibly the clearest signature of their past
history and future significance.  Eventually, the main sequence turnoff will
reach down to their mass, exposing those Li-depleted stars as canonical blue
stragglers when normal stars of that mass evolve away.  Arguing {\it against}
this unified view is the observation that the three Li-depleted stars at
$T_{\rm eff}~\simeq~6300$~K are {\it all} binaries, whereas very few of the
cooler systems show evidence for binarity; it is thus possible that two
separate mechanisms are responsible for the production of Li-deficient
main-sequence halo stars.
\end{abstract}


\vfill
\eject

\section{Introduction}					

$^7$Li is destroyed in stellar interiors where temperatures exceed
$2.5\times 10^6$~K, and Li-depleted material can in principle reach the
stellar surfaces where it can be observed.  Thus, if one is to infer
pre-stellar $^7$Li abundances from current-epoch observations, it is important
to understand the stellar processing of this species.
It has widely, though not universally, been supposed that warm ($T_{\rm
eff}~>~5700$~K), metal-poor ([Fe/H]~$<~-1$) stars retain their pre-stellar
abundances (Spite \& Spite 1982; Bonifacio \& Molaro et al. 1997; but see also
Deliyannis 1995; Ryan et al.  1996).  Although claims had been made of an
intrinsic spread in the Li abundances by 0.04 -- 0.1 dex (Deliyannis,
Pinsonneault, \& Duncan 1993; Thorburn 1994), Ryan, Norris \& Beers (1999)
attributed these to an embedded $A$(Li) vs [Fe/H] dependence , and
underestimated errors, respectively.  Ryan et al. (1999) set tight limits
on the intrinsic spread of $^7$Li in metal-poor field stars
as essentially zero, stated conservatively as
$\sigma_{\rm int} < 0.02$~dex.  However, the subset of ultra-Li-deficient stars
identified by Spite, Maillard, \& Spite (1984), Hobbs
\& Mathieu (1991), Hobbs, Welty \& Thorburn (1991), Thorburn (1992), and Spite
et al. (1993) stands out as a particular exceptional counter-example to the
general result.  These stars have only upper limits on their $^7$Li abundances,
typically 0.5~dex or more below otherwise similar stars of the same $T_{\rm
eff}$ and metallicity.  Detailed studies of other elements in these objects
have revealed some chemical anomalies, but none common to
all, or which might explain {\it why} their Li abundances differ so clearly
from those of otherwise similar stars (Norris et al. 1997a; Ryan, Norris \&
Beers 1998).

In contrast to the situation for Population II stars, a wider range of Li
behaviors is seen in Population I. In addition to a stronger increase with
metallicity, thought to be due to the greater production of Li in later phases
of Galactic chemical evolution (Ryan et al. 2001), there is also substantial
evidence of Li depletion in certain temperature ranges. Open cluster
observations, for example, show steep dependences on temperature for $T_{\rm
eff} \la 6000$~K (e.g., Hobbs \& Pilachowski 1988) and in the region of the
F-star Li gap (6400~K~$<~T_{\rm eff}~<$~7000~K; Boesgaard \& Tripicco 1986).
More problematic, for the young cluster $\alpha$~Per (age 50 Myr) and the
Pleiades (age 100 Myr), is the presence of a large apparent Li spread even at a
given mass. Various explanations have been proposed involving mixing in
addition to that due to convection. Extra mixing processes include
rotationally-induced mixing 
(e.g., Chaboyer, Demarque \& Pinsonneault 1995), structural changes associated
with rapid rotation (Mart\'in \& Claret 1996), and different degrees of
suppression of mixing by dynamo-induced magnetic fields (Ventura et al. 1998).
Gravity waves have been proposed as yet another different mixing mechanism
(Schatzman 1993; Montalb\'an \& Schatzman 1996).  
Consensus has not yet emerged concerning the range of possible mechanisms, or
the relative importance of each.  Jeffries (1999) even questions the reality of
a Li abundance spread in low mass Pleiades stars, 
due to a similar spread being seen in the
\ion{K}{1} resonance line. Amongst older open clusters, the spread at a given
effective temperature is generally much less, though M67 
(Jones, Fischer, \& Soderblom 1999) is an exception. A class of stars
with higher
lithium abundances than otherwise similar stars is
short-period tidally-locked binaries (Deliyannis et al. 1994; Ryan \&
Deliyannis 1995) which give credence to the view that physics related to
stellar rotation can and does influence the evolution of Li in approximately
solar-mass stars.

The fraction of warm, metal-poor stars that fall in the ultra-Li-deficient
category has previously been estimated at approximately 5\% (Thorburn 1994).
However, recent measurements of Li in a sample of 18 warm ($T_{\rm
eff}~\ga~6000$~K), metal-poor ($-2~\la~$[Fe/H]~$\la~-1$) stars yielded four
ultra-Li-deficient objects, i.e. more than 20\% of the sample (Ryan et al.
2001).  The Poisson probability of a 5\% population yielding 4 or more objects
in a sample of this size is just 0.013.  Clearly, the selection criteria for
this sample have opened up a regime rich in ultra-Li-poor stars.  We now
examine those criteria, and discuss the implications for the origin of such
systems and for our understanding of Li-poor and Li-normal stars.

We note some similarities between Li-deficient halo stars and blue stragglers.
Although these two groups have traditionally been separated due to the
different circumstances of their {\it discovery}, we question whether there is
a reliable {\it astrophysical} basis for this separation.  One must ask whether
the process(es) that gives rise to blue stragglers is capable only of producing
stars whose mass is greater than that of the main sequence turnoff of a
$\sim$13 Gyr old population. If, as we think is reasonable, the answer is
``no'', then one may ask what the sub-turnoff mass products of this process(es)
would be. Our proposal is that they would be Li-deficient, but otherwise
difficult to distinguish from the general population, and in this regard very
similar to the ultra-Li-deficient halo stars.

\section {Observations of the Ultra-Li-Poor Halo Stars} 

The ultra-Li-poor halo stars we consider were identified serendipitously in a
study of predominantly high proper-motion halo stars having $T_{\rm
eff}~^>_\sim~6000$~K and $-2~^<_\sim$~[Fe/H]~$^<_\sim~-1$, and are listed in
Table~1(a).  Details of the sample selection and abundance analysis are given
by Ryan et al. (2001); the key points are that high resolving power
($\lambda/\Delta\lambda~\simeq~50000$) \'echelle spectra were obtained,
equivalent widths were measured, and abundances were computed using a model
atmosphere spectrum-synthesis approach.  Two of the Li-poor stars were
subsequently found to have previous Li measurements; Wolf~550 was identified as
G66-30, and G202-65 had been observed by Hobbs \& Mathieu (1991) in a study
targeted at blue stragglers.  The new spectra of the four stars, plus one with
normal Li for comparison, are shown in Figure~1. The comparison star,
CD$-31^\circ$305, has $T_{\rm eff}~=~5970$~K, [Fe/H]~=~$-1.0$, and
$A$(Li)~=~2.24 (Ryan et al. 2001).  For convenience, previously known
Li-depleted halo stars are listed in Table~1(b).  The full sample of Ryan et
al. (2001) is plotted in Figure~2, along with additional stars from the
literature.

It is immediately apparent that three of the four ultra-Li-deficient stars are
amongst the hottest in our sample, though not {\it the} hottest in the figure.
It seems likely that high temperature is one biasing characteristic of these
objects.  The stars with $T_{\rm eff}~>~6300$~K and {\it normal} Li abundances
are listed in Table~1(c).  These have had comparatively high values of
de-reddening applied, and it is possible that they are in reality cooler than
Table~1 shows. An indication that high temperature is not the {\it only}
biasing characteristic of ultra-Li-poor stars is that the Ryan et al. (1999)
study of 23 very metal-poor ($-3.5~^<_\sim$~[Fe/H]~$^<_\sim~-2.5$) stars
in the same temperature range included only one ultra-Li-deficient star,
G186-26. This rate, 1 in 23, is consistent with previous estimates for
Population II stars as a whole.  However, very few relatively metal-rich
($-2~^<_\sim $~[Fe/H]~$^<_\sim~-1$) halo stars in this temperature range had
been studied previously, so earlier works may have been biased against
discovering ultra-Li-poor objects.  It appears, then, that the fraction of
ultra-Li-deficient stars is higher as metallicity increases.  This may explain
why our study, which targeted stars in the higher metallicity range {\it and}
with $T_{\rm eff}~>~6000$~K, was so successful at yielding ultra-Li-deficient
stars.  Figure~3 shows the distribution of objects in the $T_{\rm eff}$,~[Fe/H]
plane.  Whereas previously no region of parameter space stood out as
``preferred'' by Li-deficient stars, the objects are now more conspicuous as a
result of their high temperatures and relatively high metallicities.

Also shown in Figure~3 are the $T_{\rm eff}$ of the main-sequence turnoff as a
function of metallicity, for 14 and 18~Gyr isochrones. The isochrones shown are
the oxygen-enhanced curves of Bergbusch \& VandenBerg (1992; solid curves;
Y=0.235), and, for comparison, the Revised Yale Isochrones of Green, Demarque
\& King (1987; dotted curves; Y=0.24). Clearly there is disagreement of
$\simeq$4~Gyr between the two sets as to the ages that would be assigned to
these stars, and there are uncertainties in the color-$T_{\rm eff}$
transformations that have been applied to the observed data. However, these
difficulties are not the issue here. Rather, we use the isochrones to indicate
the {\it shape} of the turnoff locus in the $T_{\rm eff}$ vs [Fe/H] plane, and
on that point the four isochrones are in overall agreement. They emphasise that
even though HD~97916 is cooler than five other Li-depleted stars in the study,
it is nevertheless close to the turnoff. That is, a star with $T_{\rm eff}$ =
6100~K would appear below the turnoff if [Fe/H] = $-3$, but will be close to
the turnoff if [Fe/H] = $-1$. Even excluding the definite blue straggler
BD+25$^\circ$1981, there are four Li-depleted stars amongst the eight whose
symbols lie above or touch the 14~Gyr Revised Yale Isochrone. Clearly, all of
these are very close to the turnoff once their metallicities are taken into
account.\footnote{ We resist the temptation to speak of a {\it single} locus
for the turnoff because of the possibility that an age spread exists at a given
metallicity.  That issue has not yet been settled for the globular clusters
(see Piotto et al. 2000 and Chaboyer 2000), despite those systems being better
constrained. For the same reason, and because of random errors in the effective
temperature estimates, we refrain from debating whether a particular star lying
close to the turnoff is definitely above or below the turnoff.}
Besides these Li-depleted stars close to the turnoff, four are 100--200~K
cooler than the turnoff. We discuss later in this paper whether the these two
groupings might have different origins.

\section {Traditional blue stragglers} 

Blue stragglers are recognised observationally as stars that are considerably
bluer than the main-sequence turnoff of the population to which they belong,
but having a luminosity consistent with main-sequence membership.  Such objects
were originally identified in globular clusters (e.g., M3; Sandage 1953), but
are also known in the field (e.g., Carney \& Peterson 1981), and in Population
I as well as Population II (e.g., Leonard 1989; Stryker 1993). Their origin is
not known with certainty, and it is possible that more than one mechanism is
responsible for their presence.  A range of explanations was examined by
Leonard (1989), but the discovery of Li destruction in blue stragglers in the
halo field and the open cluster M67 led Hobbs \& Mathieu (1991) and Pritchet \&
Glaspey (1991) to conclude that ``virtually all mechanisms for the production
of blue stragglers {\it other than} mixing, binary mass transfer, or binary
coalescence appear to be ruled out ... .'' As Hobbs \& Mathieu emphasized,
internal mixing alone is also ruled out; mixing out to the surface is required.

Recent advances in high-resolution imaging have verified that the blue
straggler fractions in at least some globular clusters are higher in their
cores, strongly supporting the view that some blue stragglers are formed
through stellar collisions, probably involving the coalescence of binary stars
formed and/or hardened through exchanges, in these dense stellar environments
(e.g., Ferraro et al. 1999).  However, it is neither established nor required
that a single mechanism will explain all blue stragglers, and it is unclear how
the field examples and those in the tenuous dwarf galaxy Ursa Minor (Feltzing
2000, priv. comm.) relate to those in the dense cores of globular clusters.  Probably even
the halo field and dwarf spheroidal stars formed in clusters of some
description (since the formation of stars in isolation is unlikely), but one
should not be too quick to link the properties of surviving globular clusters
to diffuse populations. This view is supported by Preston \& Sneden's (2000)
conclusion that at more than half (62\% -- 100\%) of their field blue
metal-poor binaries are blue stragglers formed by mass transfer rather than
mergers, due to the long orbital periods and low eccentricities of the field
systems they observed.  Their conclusion is entirely consistent with the views
of Ferraro et al. (1995), who ascribed blue straggler formation to interactions
{\it between} systems in high-density environments, but {\it within} systems
(primordial binaries) in lower-density clusters.  In contrast to but not
contradicting Preston \& Sneden's result for field systems, Mateo et
al.  (1990) argue that all of the blue stragglers in the globular cluster
NGC~5466 are the result of close binary mergers.

The mechanism for Li destruction in field blue stragglers is not known.  It is
unclear what degree of mixing will occur as a result of coalescence. Early work
by Webbink (1976) suggested substantial mixing would occur, whereas more recent
simulations of head-on collisions by Sills et al. (1997), and grazing
collisions and binary mergers by Sandquist, Bolte, \& Hernquist (1997), have
suggested otherwise. Sills, Bailyn \& Demarque (1995) argue, however, that to
account for the blue stragglers observed in NGC~6397, mixing is nevertheless
required (unless the collision products have more than twice the turnoff mass),
and may occur after the initial coalescence. This is perhaps consistent with
the result of Lombardi, Rasio, \& Shapiro (1996) that some mixing could occur
as a merger remnant re-contracts to the main sequence.  Due to the fragility of
Li, if some mixing of surface material does occur during the coalescence it
will at least dilute, and possibly also destroy, any lithium remaining in the
stars' thin convective surface zones up to that time. One might suppose that
mass transfer in a detached system also destroys Li, though one could also
imagine gentle mass-transfer processes where the rate is slow enough that the
original envelope is not subjected to additional mixing, and where the
transferred matter itself does not undergo additional Li-destruction.  Of
course, mass transfer via Roche lobe overflow in a detached system, or wind
accretion from a more distant companion, involve mass from an evolved star
which may have {\it already} depleted its surface Li due to single-star
evolutionary processes.  Consequently, the mass transferred may be already
devoid of Li, as in the scenario quantified by Norris et al. (1997a).

We also note the possibility that the accretor in a mass-transfer system, or the
progenitors of a coalescence, was (were) devoid of Li prior to that event. Li is
(normally) preserved in halo stars only over the temperature range from the 
turnoff ($T_{\rm eff}~\simeq~6300$~K) to about $T_{\rm eff}~\simeq~5600$~K, 
corresponding to a mass range from 0.80 down to 0.70~$M_\odot$. Therefore it is 
likely that any mass accretor, and certain that any merger remnant, now seen in 
this mass range began life as one (or two) stars with initial mass(es) less 
than 0.70~$M_\odot$ and had already destroyed Li normally, as lower-mass
stars are known to do, prior to mass exchange. In such a scenario, it is not
{\it necessary} for any Li to have been destroyed as a result of the 
blue-straggler formation process itself, though this could occur as well.

\section {Discussion} 

In view of the distributions of the ultra-Li-deficient stars in the 
$T_{\rm eff}$,~[Fe/H] plane, with four at the turnoff and four 100--200~K
cooler, we consider whether all represent the 
same phenomenon, or the possibility that two distinct processes have been
in operation.  It is not a trivial matter to answer this question, because we
do not know with certainty what mechanism(s) has affected any of the stars.
However, we explore a number of possibilities in the discussion that follows.
Ignoring again the obvious blue straggler BD+25$^\circ$1981, of the 111 stars
shown in Figure~3, 8 are ultra-Li-deficient.  If all ultra-Li-poor stars have
the same origin, then we should begin by restating the frequency of such
Li-weak objects as $\simeq$7\% of plateau stars rather than $\simeq$5\% as
estimated previously when the parameter space was incompletely sampled, and
with strong metallicity and temperature dependences in that fraction.

\subsection {Do Ultra-Li-Deficient Stars and Field Blue Stragglers Share a 
Common Origin?}

Historically, blue stragglers and ultra-Li-deficient stars have been regarded
as separate phenomena. However, we have been driven to consider whether there
is any astrophysical basis for this separation.  One must ask whether the
process(es) that gives rise to blue stragglers is capable only of producing
stars whose mass is greater than that of the main sequence turnoff of a
$\sim$13 Gyr old population. If, as we think is reasonable, the answer is
``no'', then one may ask what the sub-turnoff mass products of this process(es)
would be. Our proposal is that they would be Li-deficient, but otherwise
difficult to distinguish from the general population.\footnote{ The likelihood
of sub-turnoff mass objects being produced by the blue-straggler forming
process is independently addressed in the model by Preston \& Sneden (2000, \S
5.3), which came to our attention during finalisation of this manuscript.}

For ultra-Li-poor stars redder than the main sequence turnoff, Hipparcos
parallaxes have established that G186-26 is on the main sequence rather than
on the subgiant branch. Of those {\it at} the turnoff, Wolf~550, G202-65, and
BD+51$^\circ$1817 also have Hipparcos parallaxes; two are almost certainly
dwarfs, while G202-65 is subject to larger uncertainties and may be more
evolved (see Ryan et al. 2001, Table~2). The argument that the
evolutionary rate of subgiants is too rapid to explain the high frequency of
observed Li-deficient objects, which persuaded Norris et al.
(1997a) to reject the proposition that they might be the {\it redward}-evolving
(post-turnoff) progeny of blue-stragglers, is therefore redundant. 
However, the detection of several
Li-weak stars at the bluest edge of the colour distribution has prompted us to
re-examine their possible association with blue stragglers.

We would describe G202-65 as ``at'' the turnoff rather than classify it as a
blue straggler in the conventional sense, as it is only marginally hotter
(bluer) than the main sequence turnoff for its metallicity (see Figure~3). 
Hobbs \& Mathieu, on the other hand, classified it as a
blue straggler, based presumably on the photometry of Laird, Carney \& Latham
(1988) which they referenced. (Indeed, Carney et al (1994) declare it as a
``blue straggler candidate'', and Carney et al. (2000) treat it as one, though
acknowledging at the same time that some normal stars may be included in this
classification.) Our purpose is {\it not} to 
debate how this star should be classified, but rather to underline the main
suggestion of our work, that the blue straggler and halo ultra-Li-deficient
stars may have a common origin.  Although blue
stragglers have historically been recognised because they are bluer than the
main-sequence turnoff, it is essential to remember that stars that have accreted
mass from a companion, or that result from a
coalescence can have a mass less than the current turnoff.  Such stars would
be expected to share many of the properties of blue stragglers, but would not
{\it yet} appear bluer than the turnoff. However, at some future time, once the
main-sequence turnoff reaches lower masses, these non-standard objects would
lag the evolution of normal stars and hence appear bluer, showing canonical
straggling behaviour.  Therefore, such stars might, for the present, be
regarded as ``blue-stragglers-to-be,''\footnote{Independently, Carney et al. (2000) 
have noted this possibility, and models by Portegies Zwart (2000)
predict the existence of such objects.} and our speculation is that
the ultra-Li-deficient halo stars in are in fact members of such a population.
Note that this proposition is distinct from that of {\it redward}-evolving
systems considered and rejected by Norris et al.  (1997a).

If ultra-Li-deficient stars and blue stragglers are manifestations of the same
process, then Li deficiency may be the only way of distinguishing
sub-turnoff-mass blue-stragglers-to-be from normal main-sequence stars, prior to
their becoming classical blue stragglers.  Mass transfer during their
formation may also help clarify some of the unusual element
abundances found by Norris et al. (1997a; see also Ryan et al. 1998).  Whereas
an appeal to extra mixing (in a single-star framework) to explain the Li
depletion would not necessarily affect other elements, mass transfer in a
binary with an AGB donor may be capable of
altering s-process abundances as well.  In this regard, we recall that two of
the ultra-Li-deficient stars studied by 
Norris et al. (1997a; also Ryan et al. 1998) had non-standard Sr and Ba
abundances. Mass transfer from an RGB donor would presumably leave a different 
chemical signature.\footnote{Amongst very metal-poor stars with [Fe/H] $<$ $-2.5$,
as many as 25\% have C overabundances (e.g. Norris, Ryan, \& Beers 1997b).
At least some but not all of these (Norris, Ryan, \& Beers 1997c) have
s-process anomalies. Detailed studies have yet to be completed,
so it is unclear what fraction of stars are formed from anomalous
material and what fraction became modified later in their life.
Whilst we cannot presently rule out the possibility that the s-process anomalies
seen in some ultra-Li-deficient stars were inherited at birth, our expectation
is that mass transfer from a companion star will be a more common mechanism.}

Some constraints on the progenitors of the Li-deficient stars may be obtained
from their rotation rates and radial velocity variations.  Webbink's (1976)
calculations of a coalesced star ($M_{\rm total}~=~1.85~M_\odot$) show that a
high rotation rate is maintained at least until it reaches the giant branch.
In contrast, previously known blue stragglers appear not to have uncommonly
high rotation rates (e.g., Carney \& Peterson 1981; Pritchet \& Glaspey 1991).
This tends to argue against the blue stragglers 
as having originated from coalesced
main-sequence contact binaries, and points towards one of the other binary
mass-transfer scenarios, unless mass loss (e.g., via Webbink's excretion disk)
and magnetic breaking can dissipate envelope angular momentum during the main
sequence lifetime of a coalesced star. To spin down, stars must have a
way of losing surface angular momentum. In single stars, most of this is 
believed to occur during the pre- and early-main-sequence phase when magnetic
coupling of the stellar surface to surrounding dust creates a decelerating
torque on the star. It is not clear that two mature stars which merge will
still have this coupling, because of the much lower mass loss rates beyond the
early stages of evolution (unless they produce an excretion disk) and lower
magnetic field strengths. (See also discussion by Sills et al. 1997, \S5.5.)
Leonard \& Livio (1995) have proposed that the merger product acquires the
distended form of a pre-main-sequence-like star which then spins down as it
again approaches the main sequence, losing angular momentum in much the same
way as conventional pre-main-sequence stars.  
\footnote{Although stellar collisions
will be rare for stars in the field, we should recall that most stars are
probably born in clusters, and prior to cluster dissolution, collisions would
have greater probability.}

For the four stars observed in this work, three had previous radial velocity
measurements accurate to $\simeq$~1~km~s$^{-1}$ (Carney et al. 1994). The new
measurements (Ryan et al. 2001; Table 2) showed residuals of +1.0
(BD+51$^\circ$1817), $-$3.3 (G202-65), and $-$6.9~km~s$^{-1}$ (Wolf~550);
compared with the expected radial velocity accuracy of
$\sigma_v$~=~0.3--0.7~km~s$^{-1}$, these are consistent with significant
motion. Carney et al. (2000) indicate periods of 168 to 694~days for these
systems, and low eccentricities, except for Wolf~550 ($e$~=~0.3).  Similarly,
the metal-poor field blue straggler CS~22966-043 has an orbital period of
319~days (Preston \& Landolt 1999). If the brighter component has a mass of
0.8~M$_\odot$ and its companion has a mass between 0.4 and 1.4~M$_\odot$
(appropriate to a white dwarf) then the {\it current} semi-major axis of the
system will be in the range $a$~=~200--260~R$_\odot$ (from Kepler's Third
Law).\footnote{Carney et al. (2000) argue that all of their blue-straggler
observations are consistent with 0.55~M$_\odot$ companions having a canonical
white-dwarf mass.} Their second system, CS~29499-057, may have an even longer
period of 2750~days, implying $a$~=~900--1100~R$_\odot$. The periods of these
and Carney et al's systems, and hence their large current separations, are more
compatible with mass loss from an evolved companion rather than being
short-period systems in contact on the main sequence.

The evidence presented to date has argued against internal mixing alone as an
adequate explanation for the ultra-Li-deficient stars whose neutron-capture
elements show abundance anomalies. Note, though, that certainly not all
ultra-Li-deficient stars and blue stragglers exhibit neutron-capture element
anomalies (Carney \& Peterson 1981; Norris et al. 1997a; Ryan et al. 1998).  If
mass transfer has occurred, systems in which s-process elements are abnormal
would presumably indicate material originating with an AGB companion, whereas
s-process-normal remnants would indicate mass transfer during an earlier stage
of evolution (RGB) or from a pre-thermal-pulsing AGB mass donor. (We have no
data on the N abundance, and the CH band in these stars is too weak to hope to
measure the $^{12}$C/$^{13}$C ratio.) Likewise, the rotation rates of both blue
stragglers and ultra-Li-deficient stars are apparently normal, arguing against
coalescences having already occurred on the main sequence. Of the three
mechanisms found to be viable by Pritchet \& Glaspey (1991) and Hobbs \&
Mathieu (1991), this leaves mass transfer from a companion as the only one
remaining, {\it if} we are correct in speculating that the ultra-Li-deficient
and blue straggler phenomena are manifestations of the same process.

\subsection {The Hot Stars in Isolation} 

In the absence of an adequate theory for why eight otherwise-normal halo stars
(excluding the traditional blue straggler BD+21$^\circ$1981) should have low
(zero?) Li abundances, it may be useful to consider the hot subsample
(6200~K~$^<_\sim$~$T_{\rm eff}$~$^<_\sim$~6300~K) as a distinct group. Several
possibilities then arise that might account for the observed Li deficiency,
including diffusion (the sinking of Li to below the photosphere), the F-star Li
dip, and an unknown process that may be responsible for depletion in some (but
not all) disk stars.  We consider each of these in turn.  We note that the
three Li-deficient stars with $T_{\rm eff}~\simeq~6300$~K are confirmed
binaries, whereas most cooler ones show no evidence of binary motion.  The
binary/single distinction between warmer/cooler Li-depleted stars is
pronounced; see Table~1, where the binary status (Carney et al. 1994, 2000;
Latham 2000, priv.comm.) is given in the final column. If such a dichotomy is
maintained as more Li-poor systems are discovered, it may indicate a genuine
difference in the origin of the turnoff and sub-turnoff systems.

\subsubsection{Diffusion} 

Deliyannis, Demarque \& Kawaler (1990) and Proffitt \& Michaud (1991) have
computed the predicted effects of diffusion on the surface Li abundances
of warm halo stars.  Diffusion is more significant in hotter stars because
their surface convective zone is thinner.  The degree of depletion expected at
$T_{\rm eff}~\sim~6300$~K is a function of effective temperature,
changing by $\simeq$~0.2~dex per 100~K in the former (for $\alpha$~=~1.1), and
$\simeq$~0.2 and $>~0.2$~dex per 100~K in the latter (for $\alpha$~=~1.7 and
1.5 respectively). This does not match the behavior observed (see Figure 2).
For comparison, our ultra-Li-poor stars are depleted by $^>_\sim0.8$~dex. This
alone appears to rule out diffusion as the explanation, except possibly for the
lower-$\alpha$ model of Proffitt \& Michaud. However, Li diffusion appears to
have been inhibited in all other metal-poor samples (e.g., Ryan et al. 1996),
so it would be unusual to see it suddenly present and with such effect only in
isolated stars in our new sample.

\subsubsection {The F-Star Li Dip} 

Boesgaard \& Tripicco (1986) and Hobbs \& Pilachowski (1988) showed that Li is
severely depleted in Population I open cluster stars over the interval
6400~K~$<~T_{\rm eff}~<~7000$~K. Various explanations have been proposed,
including mass loss (e.g., Schramm, Steigman, \& Dearborn 1990), diffusion
(e.g., Turcotte, Richer \& Michaud 1998), and slow mixing of various forms
(e.g., Deliyannis \& Pinsonneault 1997), but none has been convincingly
established as responsible, and several mechanisms may be acting in concert
(e.g., Turcotte et al.).  Whatever the correct explanation(s), is it possible
that the hottest ultra-Li-deficient stars are encroaching on this regime and
are affected by this phenomenon?  Although this cannot be ruled out completely
for the hot subset, especially since we have questioned the reliability of the
$E(B-V)$ (and hence $T_{\rm eff}$) values of the hottest Li-preserving stars in
Figure~2, the onset of destruction in the F-star dip seems too gradual with
$T_{\rm eff}$ to explain the new data. The Hyades observations (Boesgaard \&
Tripicco 1986) show a decrease of only 0.3~dex from 6200 to 6400~K,
substantially less than the $^>_\sim 0.8$~dex deficit in the ultra-metal-poor
objects around 6300~K.\footnote{The critic could object that there are
deficiencies in comparing metal-rich and metal-poor objects in this fashion. We
would agree, but would also note that such a comparison is justifiable if only
to show that the two behaviors are dissimilar.} As noted above, Hipparcos
parallaxes are available for five of the eight known ultra-Li-deficient stars
and, with the possible exception of G202-65, rule out the possibility that
these stars are redward-evolving {\it descendants} of the Li-dip.

\subsection {Anomalously-Li-Depleted Disk Stars}

Lambert, Heath \& Edvardsson (1991) found that, in almost all cases, the low Li
abundances in their Population~I sample could be ascribed to their being
evolved descendants of Li-dip stars, or else being dwarfs exhibiting the
Li-depletion that increases towards {\it lower} temperature, as is normally
associated with pre-main-sequence and/or main-sequence burning.  Anomalously
high Li depletions were found in only 1--3 cases out of some 26 old-disk stars,
and for a similar fraction of young-disk stars.  Based on this
fraction, Lambert et al. proposed that a new class of highly Li-depleted stars,
comprising less than about 10\% of the population, might exist.  It is
interesting to note that this proposal pre-dated the discovery of
ultra-Li-deficient halo dwarfs.

The uncertain number of cases stated above arises because Lambert et
al. recognised that uncertainties in the stellar luminosities, and hence mass,
could drive stars into or out of the region of importance.  We now have the
benefit of accurate Hipparcos parallaxes.  These indicate that two of the seven
stars highlighted by their study, HD~219476 and HR~4285,  are indeed
considerably more massive than reported in Lambert et al.'s tables and hence
are probably descendants of the Li gap, thus reducing the number of {\it
genuine} cases to 2 out of 26 old-disk stars, and 3 out of a similar number of
young-disk stars.  That is, the fraction of anomalously Li-depleted stars
appears to be around 8-10\%, albeit sensitive to small-number statistics.
\footnote{Errors in temperature could reduce these cases further.}
Ultra-Li-depleted Population I stars are also seen in young open clusters. 
They can be recognised, for example, in Fig.~1 of
Ryan \& Deliyannis (1995), where $\simeq$6\% of the Hyades stars cooler
than the F-star dip  appear to be ultra-Li-deficient.

Is it possible that the Li-depleted halo stars are of the same type?  The lack
of examples in the two Pop~I and Pop~II classes to compare with precludes a
detailed analysis, but we note that we see Li deficiency in about 7\%\ of halo
objects, which is comparable to the ratio for the Pop~I objects. That is, the
Pop~I and Pop~II examples could arise due to the same process, even though it
remains unclear what that process is.  We note, for completeness, that Ryan et
al.  (2001) showed that the kinematics of the new ultra-Li-depleted stars are
clearly those of halo objects, and thus they genuinely belong to the halo
Population, despite their metallicities being close to those of the most
metal-poor thick-disk stars.

The stars remaining on Lambert et al's list of unusually Li-deficient objects
are: 
HR~3648,
HR~4657,
HR~5968,
HR~6541, and
HD~30649.
Upon searching the literature for evidence of binarity or abundance anomalies
in these systems, we found that not only was HR~4657 a 850~day period binary,
but Fuhrmann \& Bernkopf (1999) had also been driven to consider this star as a
blue straggler. It has an unexpectedly high rotational velocity (in contrast to
the blue stragglers studied by Carney \& Peterson 1981).  There is no evidence
of s-process anomalies, but other unusual characteristics of the system include
an observable soft X-ray flux and the very likely association of this object
with GRB 930131.  HR~3648 (= 16~UMa = HD~79028) is a 16.2~day period
chromospherically-active single-lined spectroscopic binary (Basri, Laurent, \&
Walter 1985).  HD~30649 (= G81-38) and HR~6541 (=HD~159332), in contrast, show
no significant evidence of binarity (Carney et al. 1994).  HR~5968 (= $\rho$
CrB) does not appear to have a stellar companion, though it has a planetary
companion (Noyes et al. 1997), but
Ryan (2000) argues that Li in this star is {\it not}
anomalous.  HR~3648 and HR~4657 have Ba abundance measurements from the study
by Chen et al. (2000).  The latter also has been observed by Fuhrmann \&
Bernkopf (1999), but neither star appears abnormal in this element.

\section {Implications and Summary} 

Ryan et al. (1999) have argued that the ultra-Li-deficient halo stars are
distinct from the majority of halo stars that occupy the Spite plateau, and, in
particular, that they do {\it not} merely represent the most extreme examples
of a {\it continuum} of Li depletion. If the association with blue stragglers
(or, for that matter, any distinct evolutionary phenomenon) is correct, then
the mechanism for their unusual abundances will at last be understood and they
will be able to be neglected with certainty from future discussion of the Spite
plateau.

In the present work, we have proposed and discussed the possibility that
ultra-Li-depleted halo stars and blue stragglers are manifestations of the same
phenomenon, and described the former as ``blue-stragglers-to-be.''  We proposed
that their Li was destroyed either during the formation process of blue
stragglers or during the {\it normal} single-star evolutionary processes of
their precursors, namely during pre-main-sequence and/or main-sequence phases
of low-mass stars, or during post-main-sequence evolution of mass donors, as in
the scenario quantified by Norris et al. (1997a).  We note that in a study
carried out separately but over the same time period as ours, Carney et al.
(2000) have examined the orbital characteristics of blue stragglers, and have
been driven towards similar considerations as we have. There are clearly still
details to be clarified, but our two groups appear to be converging on a view
unifying blue stragglers and ultra-Li-deficient systems.

Because there are numerous observational and theoretical issues surrounding
this unified view, we seek to clarify the main arguments and possibilities
using an itemised summary.

Observations:
\newline $\bullet$ 
In a study of 18 halo stars with $-2~^<_\sim$~[Fe/H]~$^<_\sim~-1$ and
6000~K~$^<_\sim~T_{\rm eff}~^<_\sim$~6400~K, we have found four 
ultra-Li-deficient objects, i.e. a 22\% detection rate.
\newline $\bullet$ 
The fraction of ultra-Li-deficient stars is very much higher amongst the
hottest and most metal-rich halo main-sequence stars ($\simeq$20\%) than
amongst cooler and more metal-poor ones ($\simeq$5\%).
\newline $\bullet$ 
Ultra Li-deficient stars exist both at the turnoff, and cooler than the
turnoff, and with well-determined main-sequence luminosities from Hipparcos.
\newline $\bullet$ 
All of the turnoff ultra-Li-deficient halo stars, but none of the sub-turnoff
ultra-Li-deficient halo stars, appear to be binaries. This may indicate that
two different mechanisms are causing the halo ultra-Li-deficient phenomenon.

Theoretical framework:
\newline $\bullet$ 
Blue stragglers may form from {\it several} mechanisms, but seem to require at
least one of either complete mixing, binary mass transfer, or
coalescence\footnote{Coalescence may be between the components of an
existing binary, possibly having been hardened via interactions with a
third star, or through direct collisions (which may also be moderated by binary
interactions).} (Hobbs \& Mathieu 1991; Pritchet \& Glaspey 1991).

Origins:
\newline $\bullet$ 
We speculate that ultra-Li-deficient stars and blue stragglers are
manifestations of the same process, and that sub-turnoff-mass
ultra-Li-deficient stars may be regarded as ``blue-stragglers-to-be.''
\newline $\bullet$ 
Li could be destroyed at several stages: 
(i) in a mass-transfer event which induces extensive mixing;
(ii) by single-star evolutionary processes (convective mixing) in a 
post-main-sequence mass donor;
(iii) by single-star evolutionary processes (mixing) in 
pre-main-sequence (or possibly main-sequence) low-mass stars prior to their
gaining mass.
\newline $\bullet$ 
Mass-transfer scenarios from an AGB star seem better able to explain the unusual
neutron-capture element ratios {\it sometimes} seen in ultra-Li-depleted stars 
(Norris et al. 1997a) than internal mixing, since
$\simeq$~0.8~M$_\odot$ core-hydrogen-burning stars are not expected to process
neutron-capture elements.  
This argues against internal mixing as the
sole explanation for the existence of ultra-Li-depleted stars with unusual
neutron-capture abundances. 
(Mass transfer from pre-AGB (most likely RGB) donors
would produce the stars with normal neutron-capture abundances.)
\newline $\bullet$ 
Coalesced binaries are expected to maintain high rotation rates until they reach
the giant branch, but neither blue stragglers nor ultra-Li-depleted halo stars
have high rotation rates. This argues against coalescence of a
binary as the explanation for these objects unless they have spun down. 
\newline $\bullet$ 
The orbital periods of metal-poor field blue stragglers 
(Preston \& Landolt 1999; Carney et al. 2000) suggest current semi-major axes 
in the range 
200--1100~R$_\odot$, arguing against these being coalescing stars (unless they 
began their lives as triple systems).
\newline $\bullet$ 
The arguments against solely internal mixing, and against coalescence of
main-sequence contact binaries, leaves mass transfer as the most viable
mechanism for field binaries.  This is {\it not} to say that Li was destroyed 
during the transfer;
it may have been destroyed by single-star mechanisms already.
\newline $\bullet$ 
The observed d$A$(Li)/d$T_{\rm eff}$ is too steep compared with models of
diffusion to be due to that process.
\newline $\bullet$ 
The observed d$A$(Li)/d$T_{\rm eff}$ is too steep compared with the Hyades
data to be due to the F-star Li dip.
\newline $\bullet$ 
The halo ultra-Li-deficient stars could be related to the Pop I
anomalously-Li-depleted stars identified in the field by Lambert et al. (1991)
and also seen in open clusters.
\newline $\bullet$ 
Hipparcos parallaxes rule out the possibility that the ultra-Li-deficient stars
are redward-evolving post-turnoff stars. They have not descended
from the F-star Li dip.

Implications:
\newline $\bullet$ 
Severe Li depletion may be the (only?) signature of sub-turnoff-mass blue 
stragglers. The halo population fraction comprising ultra-Li-poor stars is 7\%.
\newline $\bullet$ 
Understanding the ultra-Li-depleted stars as resulting from a distinct process
(not normally affecting single stars) would eliminate the need to include them
in discussions of processes affecting the evolution of normal Spite plateau
stars, and would explain why they appear so radically different from the vast
majority of halo stars (Ryan et al. 1999).

\section{Acknowledgements}

The authors gratefully acknowledge the support for this project given by the
Australian Time Assignment Committee (ATAC) and Panel for the Allocation of
Telescope Time (PATT) of the AAT and WHT respectively, and for practical
support given by the staff of these facilities.  They also express gratitude
to D. A. Latham and B. W. Carney for conveying the results of their program in 
advance of publication, and to an anonymous referee for his/her comments that 
helped us clarify our arguments. S.G.R. sends a special thanks
to colleagues at the University of Victoria: to C. J. Pritchet for a
most memorable snow-shoeing expedition on 1991 February 10 during which Li
deficiency in blue stragglers was discussed, to D. A. VandenBerg for discussing
and supplying isochrones, and to F. D. A.  Hartwick for once asking whether
there were blue stragglers in the halo field. T.C.B acknowledges partial
support from grant AST 95-29454 from the National Science Foundation.

\vfill
\eject


\clearpage
\begin{figure}[!htb]
\begin{center}
\leavevmode
\epsfxsize=160mm
\epsfbox{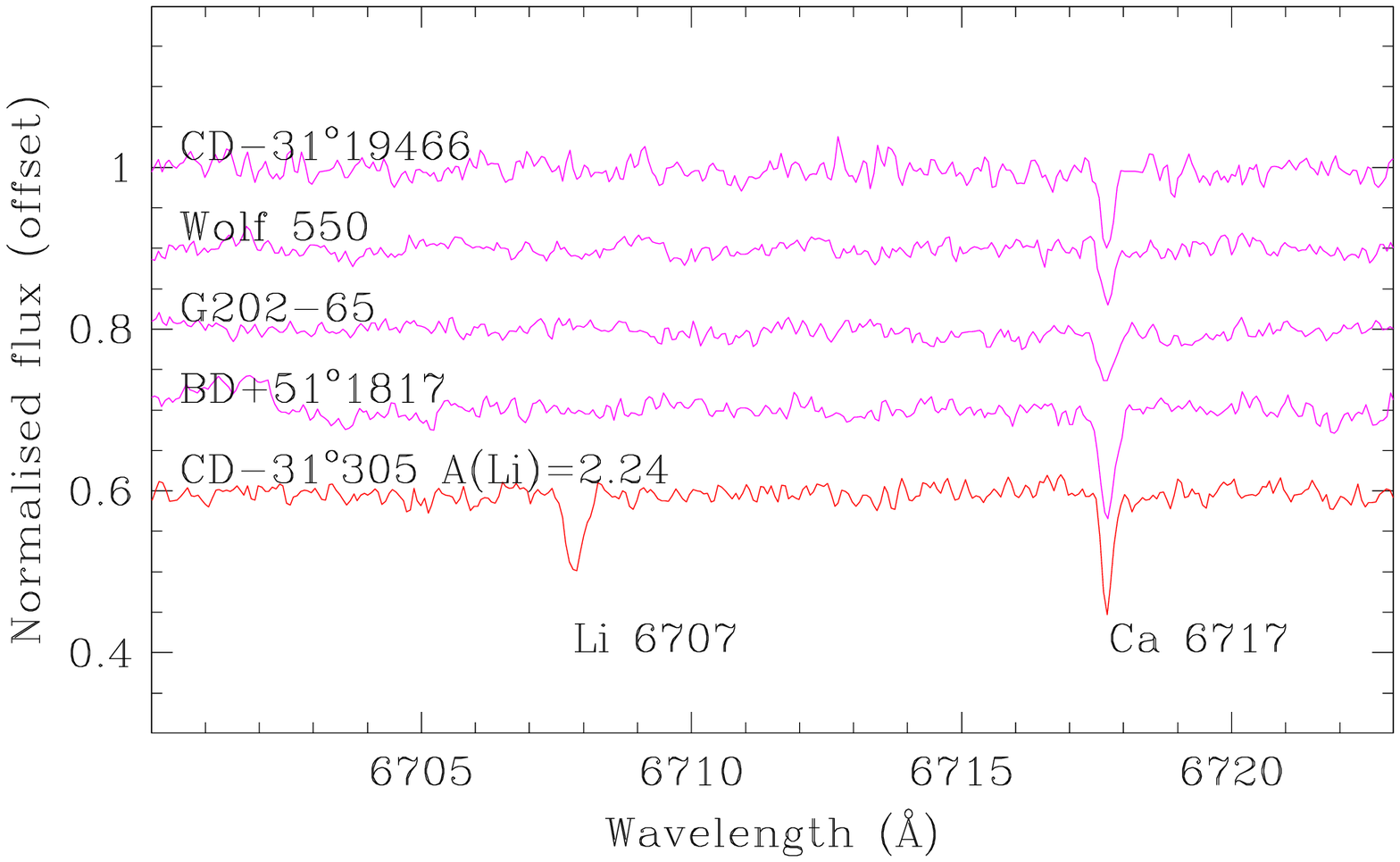}
\end{center}
\caption{
Figure 1:
Spectra in the region of the Li~6707 doublet, in order of increasing [Fe/H]. 
A fifth star, CD$-31^\circ 305$, with Li abundance close to the Spite plateau,
is shown for comparison.
}
\end{figure}

\clearpage
\begin{figure}[!htb]
\begin{center}
\leavevmode
\epsfxsize=160mm
\epsfbox{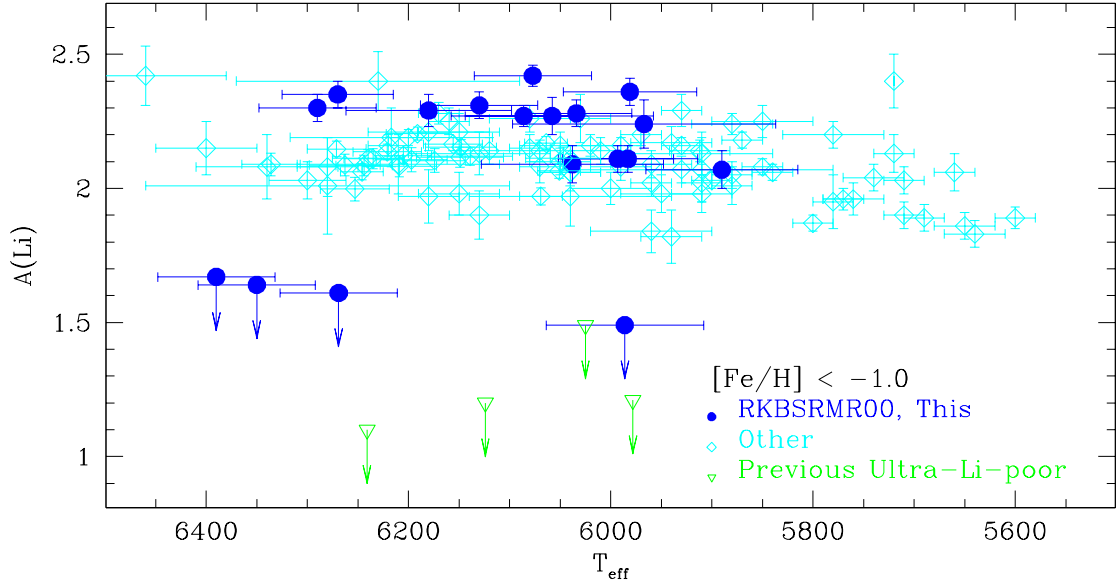}
\end{center}
\caption{
Figure 2:
Variation of $A$(Li) with $T_{\rm eff}$ for halo dwarfs with 
[Fe/H]~$<~-1.0$.  Three of the four ultra-Li-depleted stars are amongst the
hottest in the sample, even though stars down to $T_{\rm eff}~=~5900$~K were
included.  {\it Solid circles}: Ryan et al. (2000) and this study; {\it open
triangles}: previous Li-deficient observations (see Table~1); {\it open
diamonds}: data from Rebolo et al. (1988), the compilation by 
Ryan et al. (1996), Ryan et al. (1999), Norris et al. (2000), and 
Spite et al. (2000).
}
\end{figure}

\clearpage
\begin{figure}[!htb]
\begin{center}
\leavevmode
\epsfxsize=160mm
\epsfbox{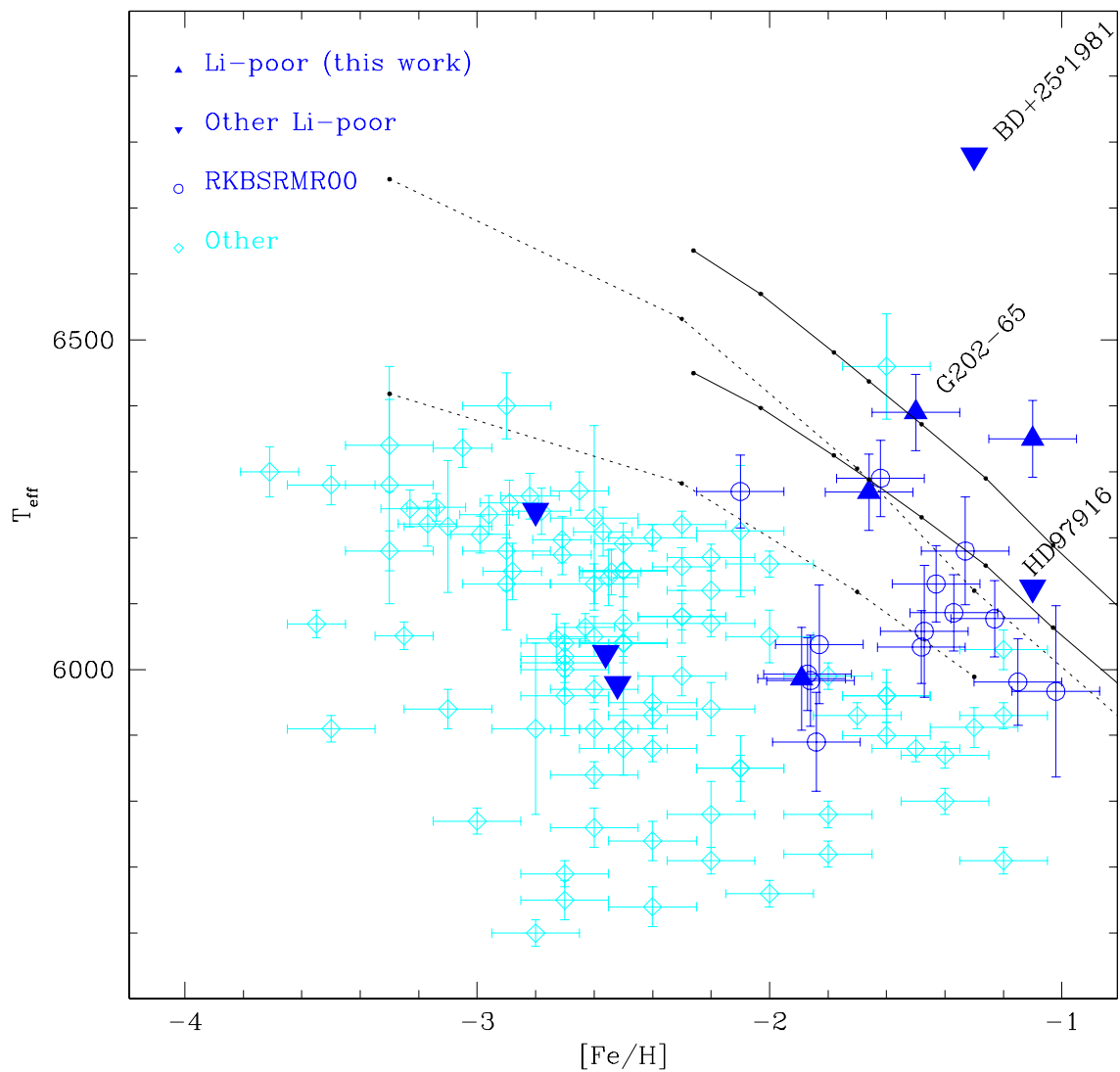}
\end{center}
\caption{
Figure 3:
Location of halo stars with known $A$(Li) in the $T_{\rm eff}$,~[Fe/H] plane.
{\it Solid triangles}: ultra-Li-deficient stars;
{\it open circles}: Ryan et al. (2001);
{\it open diamonds}: as in Figure~2.
The locus of turnoff stars of different metallicity are shown for 14~Gyr
and 18~Gyr isochrones from Bergbusch \& VandenBerg (1992; {\it solid curve})
and Green et al.  (1987; {\it dotted curve}).
}
\end{figure}

\clearpage
\begin{figure}[!htb]
\begin{center}
\leavevmode
\epsfxsize=175mm
\epsfbox{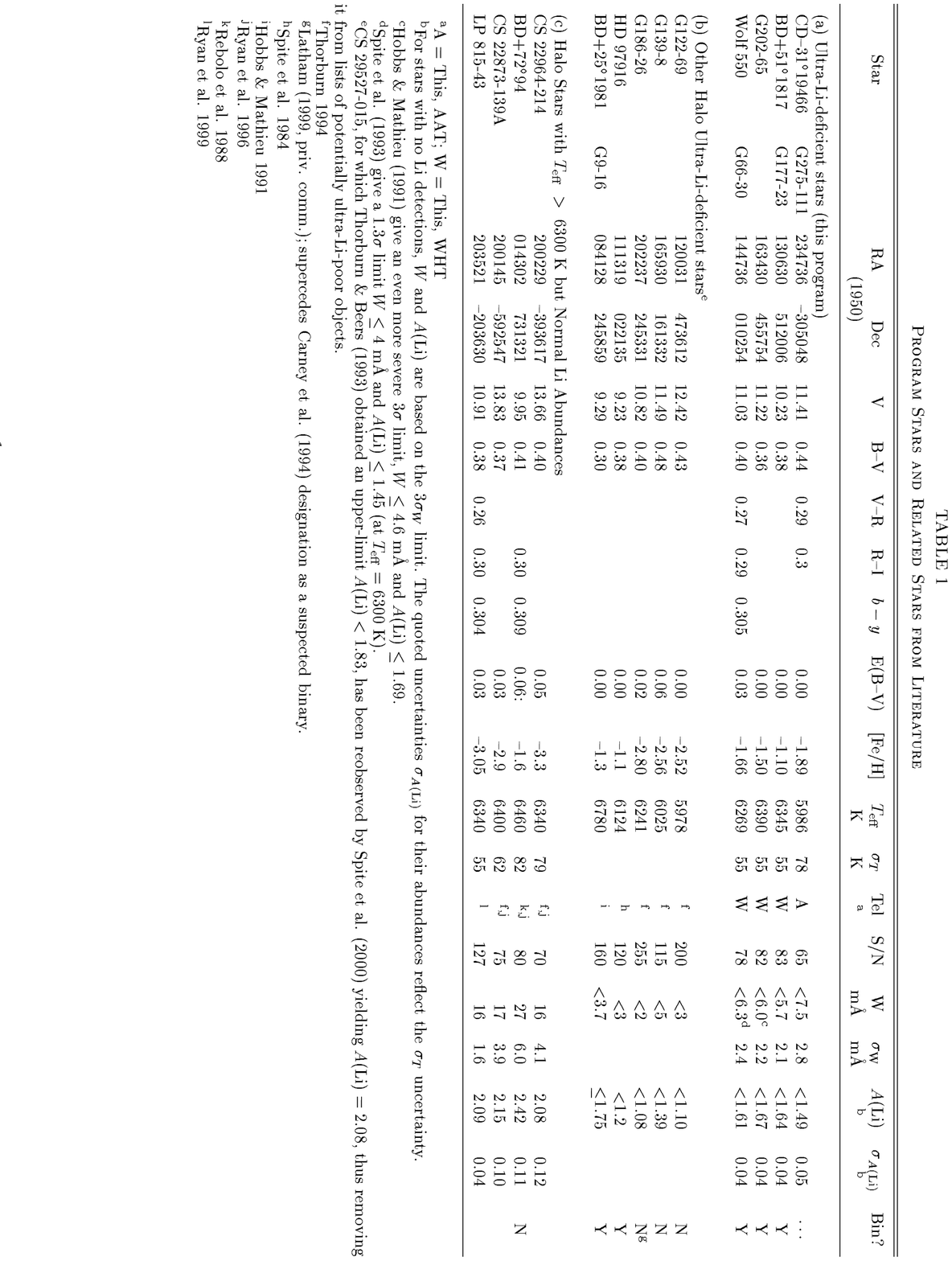}
\end{center}
\caption{Table 1}
\end{figure}

\end{document}